\documentstyle[aps,preprint]{revtex}
\begin{document}
\draft
\title{Sound propagation in density wave conductors and the effect of\\
long-range Coulomb interaction}
\author{Attila Virosztek}
\address{Research Institute for Solid State Physics, H-1525 Budapest,
P. O. Box 49, Hungary\\
and Institute of Physics, Technical University of Budapest, H-1521
Budapest, Hungary}
\author{Kazumi Maki}
\address{Department of Physics and Astronomy, University of Southern
California, Los Angeles, CA 90089-0484}
\date{\today}
\maketitle
\begin{abstract}
We study theoretically the sound propagation in charge- and spin-density
waves in the hydrodynamic regime. First, making use of the method of
comoving frame, we construct the stress tensor appropriate for quasi-one
dimensional systems within tight-binding approximation. Taking into
account the screening effect of the long-range Coulomb interaction, we
find that the increase of the sound velocity below the critical
temperature is about two orders of magnitude less for longitudinal sound
than for transverse one. It is shown that only the transverse sound wave
with displacement vector parallel to the chain direction couples to the
phason of the density wave, therefore we expect significant
electromechanical effect only in this case.
\end{abstract}
\narrowtext

\section{Introduction}

Several aspects of the collective transport associated with the phason
in charge- or spin-density waves (CDW or SDW) are still not well
understood. One of the intriguing phenomena is the electromechanical
effect observed in CDW\cite{Brill,Moz,Xi1,Xi2,Jac} and more recently in
SDW\cite{Brown}. First of all, most of the elastic moduli increase upon
entrance into the CDW or SDW state, often with a sharp dip at $T_c$, the
transition temperature. Second, some of the elastic moduli in CDW or
SDW soften when the density wave is depinned by an external electric
field in excess of the depinning threshold field $E_T$. Third, the change
in the elastic moduli due to depinning of the density wave depends on
the frequency $\omega$ of the flexural vibration\cite{Xi3}, and
decreases like $\omega^{-p}$ with $p\approx 1$. This behavior is similar
to the frequency dependence of the change in the dielectric constant
upon depinning in CDW and SDW\cite{Cava}.

We have shown earlier\cite{MV1} in the collisionless limit
that the hardening of the elastic
constants can be understood in terms of the reduction in the
quasiparticle screening of the ion potential due to the formation of
the density wave state. The electromechanical effect was interpreted
as an additional screening contribution from the collective mode of
the density wave condensate (phason) liberated by depinning. A later
extension of that theory to the experimentally more relevant
hydrodynamic limit\cite{VM1} did not modify the above picture
qualitatively. However in these papers it was assumed that the phonon
simply couples to the electronic density, and the effect of the
long-range Coulomb interaction was neglected.

The purpose of the present paper is twofold. First, we shall develop
the theory of the electron-phonon coupling for a strongly anisotropic
system, which will enable us to distinguish between the behavior of
transverse and longitudinal sound waves propagating in various
directions. Second, we shall include the effect of the long-range Coulomb
interaction following Kadanoff and Falko\cite{KF}. In Section II we
construct the electronic stress tensor (which couples to the deformation
tensor of the sound wave) for a quasi-one dimensional system following
the method of comoving frame\cite{Tsu}. In Section III we concentrate
on the quasiparticle contribution to the stress tensor correlation
functions corresponding to the pinned case. Section IV is devoted to
the examination of the coupling of the stress tensor to the phason,
which is relevant to the electromechanical effect. Our conclusions are
summarized in Section V.
A preliminary report
on this work has already been published elsewhere\cite{VM2}.

\section{Electron-phonon coupling}

Following Tsuneto\cite{Tsu} let us assume that both the ionic potential
and the electronic wavefunction are deformed in the
slowly varying sound field
${\bf u}({\bf r},t)={\bf u}\cos({\bf qr}-\omega t)$
imposed externally (extreme tight-binding limit):
\begin{eqnarray}
V({\bf r})\rightarrow&V[{\bf r}-{\bf u}({\bf r},t)]\nonumber\\
\psi({\bf r})\rightarrow&\psi[{\bf r}-{\bf u}({\bf r},t)]
(1+\nabla{\bf u})^{-1/2}.
\end{eqnarray}
This displacement is generated by the unitary operator
$U=\exp[-{\bf u}({\bf r},t)\nabla]$. Therefore the deformed
wavefunction is an eigenfunction of the transformed Hamiltonian
$h=Uh_0U^+$, where $h_0=\varepsilon(-i\nabla)$ with $\varepsilon
({\bf p})$ being the zone-periodic electronic energy spectrum.
The transformed Hamiltonian $h$ is then expanded in terms of the
deformation tensor $\nabla_iu_j$, which is much smaller than one, even
though the displacement ${\bf u}$ itself may be many times of the
lattice constant for sound propagation. We obtain $h=h_0+h_{el-ph}$,
where the Hamiltonian for the electron-phonon coupling is given by
\begin{equation}
h_{el-ph}=\sum_{i,j}(\nabla_iu_j)i\nabla_jv_i(-i\nabla),
\end{equation}
with ${\bf v}({\bf p})=\partial\varepsilon({\bf p})/\partial {\bf p}$,
the velocity of the Bloch electron.
The matrix element of $h_{el-ph}$ between Bloch states is evaluated as
\begin{equation}
\langle {\bf p}+{\bf q}|h_{el-ph}|{\bf p}\rangle =
-i\sum_{i,j}q_ju_i\tau_{ij}({\bf p}),
\end{equation}
where the stress tensor
\begin{equation}
\tau_{ij}({\bf p})=mv_i({\bf p})v_j({\bf p}),
\end{equation}
and $m$ is the bare electron mass. This expression generalizes
the stress
tensor used for an isotropic metal\cite{KF}.

In orthorombic symmetry the sound wave polarized in the ${\bf i}$
direction and propagating in the ${\bf j}$ direction couples to the
$\tau_{ij}$ component of the stress tensor, and in order to determine
the effect of that coupling on the frequency (or velocity) of the
sound, we have to evaluate the appropriate stress tensor correlation
function $\langle [\tau_{ij},\tau_{ij}]\rangle$. Once this is known,
the renormalized sound velocity can be calculated in the weak
coupling limit as
\begin{equation}
c=c_0\{1-\langle [\tau,\tau ]\rangle /2Mc_0^2\},
\end{equation}
where $c_0$ is the sound velocity without electron-phonon coupling,
$M$ is the ion mass, and for clarity we have suppressed the indeces
both for the stress tensor component and for the sound velocity.

For a highly anisotropic ($t_a\gg t_b\gg t_c$) tight-binding
dispersion
\begin{equation}
\varepsilon({\bf p})=-2t_a\cos(ap_x)-2t_b\cos(bp_y)
-2t_c\cos(cp_z)-\mu,
\end{equation}
widely used for CDW and SDW materials\cite{Yam}, the velocity and
stress tensor components are easily obtained, and their values on
the open Fermi surface can conveniently be expressed by the
component of ${\bf p}$ perpendicular to the chains (${\bf x}$
direction). However, since the Green`s functions for CDW and SDW
are usually written in the left-right spinor representation\cite{VM1},
involving measuring momenta from $\pm{\bf Q}/2$ with
the density wave wavevector ${\bf Q}=(2p_F,\pi/b,\pi/c)$,
we should express the stress tensor elements in a compatible
manner. It turns out, that for each stress tensor component the
term proportional to the unit matrix dominates, therefore we have
\begin{eqnarray}
\tau_{xx}=&mv_F^2\{1-[t_b/t_a\sin(ap_F)]^2
\sin^2(bp_y)\}\nonumber\\
\tau_{yy}=&mv_y^2[1+\cos(2bp_y)]\nonumber\\
\tau_{xy}=&mv_Fv_y\sqrt{2}\cos(bp_y).
\end{eqnarray}
Here $p_F$ is the Fermi momentum,
$v_F=2at_a\sin(ap_F)$ is the Fermi velocity in the chain
direction, while $v_y=\sqrt{2}bt_b\ll v_F$ is a typical velocity
in the perpendicular direction. In the followings we restrict
our study to the $x-y$ plane, since behavior involving the $z$
direction should be similar to that of the $y$ direction. We note,
that all components of the stress tensor depend on momentum through
the combination $\varphi=bp_y$ only.

\section{Pinned density waves}

In this section we consider sound propagation with no applied
electric field. The density wave is pinned, therefore the condensate
is unable to contribute to correlation functions, including that
for the stress tensor. Mathematically this situation can be
simulated by setting the coupling of the stress tensor
(and of the density) to the phason
to zero. Then the stress tensor couples only to the density
fluctuations, resulting in the well known Coulomb screening\cite{KF}:
\begin{equation}
\langle [\tau,\tau]\rangle=\langle[\tau,\tau]\rangle_0-
{\langle[\tau,n]\rangle_0\langle[n,\tau]\rangle_0\over
q^2/4\pi e^2+\langle[n,n]\rangle_0}.
\end{equation}
Here $n$ stands for the electronic particle density, and
$\langle[A,B]\rangle_0$ denotes a correlation function, in which
only the effect of impurity scattering is taken into account.

The density correlator $\langle[n,n]\rangle_0$ in the presence of
impurity scattering was evaluated in \cite{VM1}. A straightforward
extension of that calculation confirms that under the circumstances
of the sound experiment ($lq\ll 1$, where $l$ is the mean free path)
the stress tensor correlator has two distinct contributions:
\begin{equation}
\langle[\tau,\tau]\rangle_0=\langle\tau(\varphi)\rangle_{\varphi}^2
\langle[n,n]\rangle_0+\langle[\delta\tau(\varphi)]^2\rangle_{\varphi}
\langle[n,n]\rangle_0^{no vertex}.
\end{equation}
The first contribution features
only the average
of the stress tensor $\langle\tau(\varphi)\rangle_{\varphi}=
(2\pi)^{-1}\int_{-\pi}^\pi d\varphi\tau(\varphi)$, and is
proportional to the diffusive (vertex corrected) density correlator.
The second contribution containes only the fluctuating part
of the stress tensor component
$\delta\tau(\varphi)=\tau(\varphi)-\langle\tau(\varphi)
\rangle_{\varphi}$, and therefore it is proportional to the density
correlator $\langle[n,n]\rangle_0^{no vertex}$
calculated without vertex corrections. According to the same argument,
only the average of the stress tensor couples to the density,
therefore
\begin{equation}
\langle[\tau,n]\rangle_0=
\langle[n,\tau]\rangle_0=\langle\tau(\varphi)\rangle_{\varphi}
\langle[n,n]\rangle_0.
\end{equation}
Combining Eqs.(8)-(10) we see that in the long wavelength limit
appropriate for the sound experiment ($q\approx 1/L$, where $L$
is the sample size), the average part of the stress tensor
($s$-wave component, proportional to density) is
completely screened out by the Coulomb interaction, and only the
fluctuating part contributes to the correlation function:
\begin{equation}
\langle[\tau,\tau]\rangle=\langle[\delta\tau(\varphi)]^2
\rangle_{\varphi}\langle[n,n]\rangle_0^{no vertex}.
\end{equation}
The situation here is the same as in the electronic Raman scattering,
where the long-range Coulomb interaction suppresses the density
(charge) fluctuations, and only non $s$-wave channels
survive\cite{AG}.

The evaluation of $\langle[n,n]\rangle_0^{no vertex}$ can be done
starting with the results of \cite{VM1}. The calculation is
rather technical, therefore we delegate it to the Appendix, and
we give here the results only. Without vertex corrections there is
no diffusion pole, and both the wavenumber ${\bf q}$ and the
frequency $\omega$ could be set to zero. However we keep a finite
(but small) frequency for finite imaginary part of the correlator:
\begin{equation}
\langle[n,n]\rangle_0^{no vertex}=N_F{i\tilde\Gamma_{qp}(1-\tilde f)
\over \omega+i\tilde\Gamma_{qp}},
\end{equation}
where $N_F$ is the density of states at the Fermi surface. The
corresponding "unrenormalized" condensate density $\tilde f$
(for the general formula see Eq.(32) in the Appendix) and
quasiparticle damping $\tilde\Gamma_{qp}$ are evaluated in two
limiting cases, close to $T_c$ and close to zero temperature as
\begin{eqnarray}
\tilde f(T\rightarrow T_c)=&-2({\Delta\over 4\pi T})^2
\psi^{\prime\prime}({1\over 2}+{\Gamma\over 2\pi T})\approx
7\zeta(3)({\Delta\over 2\pi T})^2\nonumber\\
\tilde f(T\rightarrow 0)=&1-3\pi\alpha/16,
\end{eqnarray}
and
\begin{eqnarray}
\tilde\Gamma_{qp}(T\rightarrow T_c)=&2\Gamma\nonumber\\
\tilde\Gamma_{qp}(T\rightarrow 0)=&{9\pi\over 32}{\Delta
\alpha^{8/3}\over u_0^2(5-4\alpha^{2/3})}{G\over T}e^{G/T}.
\end{eqnarray}
Here $\Delta$ is the density wave order parameter, $\Gamma=
\Gamma_F+\Gamma_B/2$ is a combination of the impurity forward
and backscattering rate, $\alpha=\Gamma/\Delta$, $u_0^2=1-
\alpha^{2/3}$ and $G=\Delta u_0^3$ is the density wave gap.

As it is seen from Eq.(13), $\tilde f$ increases linearly in
$(T_c-T)$ below $T_c$, but it is slightly less than one at $T=0$
($\Gamma$ is usually an order of magnitude smaller than $T_c$).
Nevertheless, the temperature dependence of the sound velocity
in the pinned case
will still be qualitatively the same as in the collisionless
limit\cite{MV1}. The relative change of the sound velocity
compared to the normal state ($c_n$) is easily obtained from Eqs.(5),
(11) and (12) as
\begin{equation}
(c-c_n)/c_0=\lambda\tilde f,
\end{equation}
where the effective coupling
\begin{equation}
\lambda={N_F\over 2Mc_0^2}\langle[\delta\tau(\varphi)]^2
\rangle_{\varphi}.
\end{equation}
Using Eq.(7), these effective couplings for the various
sound waves are
\begin{eqnarray}
\lambda_{xx}=&{N_F(mv_F^2)^2\over 16Mc_0^2}[{t_b\over t_a\sin(
ap_F)}]^4\nonumber\\
\lambda_{yy}=&{N_F(mv_y^2)^2\over 4Mc_0^2}\nonumber\\
\lambda_{xy}=&{N_F(mv_Fv_y)^2\over 2Mc_0^2}.
\end{eqnarray}
Since $v_y/v_F\approx t_b/t_a\approx 1/10$ in many quasi one
dimensional materials, we expect that the relative increase of
the sound velocity below $T_c$ will be a factor $10^2$ smaller
for longitudinal sound than for transverse sound.

\section{Electromechanical effect}

If an external electric field in excess of the threshold field
$E_T$ of the nonlinear conductivity is applied in the chain direction,
then the condensate is depinned and is able to contribute to
various correlation functions\cite{MV1}. The best known example
is of course
the conductivity itself, but the situation is the same for the
stress tensor correlator as well. The collective contribution to
$\langle[\tau,\tau]\rangle$ can be obtained if we allow both the
stress tensor and the density to couple to the phason (in the
previous section this coupling was blocked due to pinning).In this
case the stress tensor correlator has another contribution
$\langle[\tau,\tau]\rangle^{coll}$ in addition to the one
calculated in the previous section, namely:
\begin{equation}
\langle[\tau,\tau]\rangle^{coll}={U\langle[\tau,\delta\Delta]
\rangle^{Coul}\langle[\delta\Delta,\tau]\rangle^{Coul}\over
1-U\langle[\delta\Delta,\delta\Delta]\rangle^{Coul}}.
\end{equation}
Here $U$ is the on-site Coulomb repulsion responsible for the
formation of the SDW state (for CDW it should be replaced by the
phonon propagator, but that does not affect our conclusions),
$\delta\Delta$ is the phase fluctuation of the order
parameter, and
$\langle[A,B]\rangle^{Coul}$ is the correlation function of
quantities $A$ and $B$ including the effect of the long-range
Coulomb interaction (like in Eq.(8)). As we have seen earlier, in
the long wavelength limit this yields:
\begin{equation}
\langle[A,B]\rangle^{Coul}=\langle[A,B]\rangle_0-\langle[A,n]
\rangle_0\langle[n,B]\rangle_0/\langle[n,n]\rangle_0.
\end{equation}

First we consider if the allowed coupling to the phason actually
takes place for various sound waves. According to Eq.(18) we
need to examine $\langle[\tau,\delta\Delta]\rangle^{Coul}$,
which is given by Eqs.(19) and (10) as
\begin{equation}
\langle[\tau,\delta\Delta]\rangle^{Coul}=\langle[\tau,\delta
\Delta]\rangle_0-\langle\tau(\varphi)\rangle_{\varphi}\langle
[n,\delta\Delta]\rangle_0.
\end{equation}
The density-phason correlator $\langle[n,\delta\Delta]\rangle_0$
was evaluated in \cite{VM1}. Here we only reiterate that result
in the limit of experimental interest $lq\ll c_0/v_F$ (dynamic
limit):
\begin{equation}
\langle[n,\delta\Delta]\rangle_0=iN_F{\langle\zeta(\varphi)
\rangle_{\varphi}\over 2\Delta}f_d,
\end{equation}
where in our two dimensional geometry the wavenumber ${\bf q}$
appears in $\zeta(\varphi)=v_Fq_x+\sqrt{2}v_yq_y\cos\varphi$,
and the condensate density in the dynamic limit is given by:
\begin{eqnarray}
f_d(T\rightarrow T_c)=&{\Delta^2\over 2\pi T\Gamma_B}
\psi^\prime({1\over 2}+{\Gamma\over 2\pi T})\approx
{\pi\Delta^2\over 4T\Gamma_B}\nonumber\\
f_d(T\rightarrow 0)=&1.
\end{eqnarray}
We recall that the same $f_d$ appears in the current-phason
correlator $\langle[j,\delta\Delta]\rangle_0$, as well as in
the phason propagator\cite{MV2}.
Note that $f_d$ increases from zero much faster below $T_c$
than $\tilde f$ does, and that at zero temperature it saturates
exactly to $1$. The stress tensor-phason correlator can be
calculated similarly:
\begin{equation}
\langle[\tau,\delta\Delta]\rangle_0=iN_F{\langle\tau(\varphi)
\zeta(\varphi)\rangle_{\varphi}\over 2\Delta}f_d.
\end{equation}

Now we shall examine Eq.(20) for different sound waves in order
to determine if there is a collective contribution to the
corresponding stress tensor correlator. We consider longitudinal
and transverse sound waves propagating in the ${\bf x}$ and
${\bf y}$ directions. Clearly, the second (screening) term in
Eq.(20) is nonzero only for the longitudinal sound
propagating in the chain direction (${\bf q}\parallel {\bf u}
\parallel {\bf x}$),
in which case it completely cancels the first term, leading to
no collective contribution. The other longitudinal sound
propagating perpendicular to the chains (${\bf q}\parallel
{\bf u}\parallel {\bf y}$) does not couple to the phason either,
because $\langle\tau_{yy}(\varphi)\cos\varphi\rangle_{\varphi}
=0$ (see Eq.(7)). The coupling of the transverse wave
propagating in the chain direction (${\bf q}\parallel {\bf x}$
and ${\bf u}\parallel {\bf y}$) is controlled by $\langle
\tau_{xy}(\varphi)\rangle_{\varphi}=0$, yielding again no
collective contribution. This means that in all of the above
three cases there will be no electromechanical effect.

For the rest of this section we will concentrate on the only
interesting case, when the transverse sound propagates
perpendicular to the chains (${\bf q}\parallel {\bf y}$ and
${\bf u}\parallel {\bf x}$). In this case there will be coupling
to the phason, since
\begin{equation}
\langle[\tau_{xy},\delta\Delta]\rangle^{Coul}=
\langle[\tau_{xy},\delta\Delta]\rangle_0=
iN_F{f_d\over 2\Delta}mv_Fv_y^2q_y
\end{equation}
is nonzero. Now we have to consider the denominator in Eq.(18).
Since $\langle[n,\delta\Delta]\rangle_0=0$ for $q_x=0$,
therefore $\langle[\delta\Delta,\delta\Delta]\rangle^{Coul}=
\langle[\delta\Delta,\delta\Delta]\rangle_0$, and we can use
the result for the phason propagator calculated in \cite{VM1},
which in our case reduces to
\begin{equation}
1-U\langle[\delta\Delta,\delta\Delta]\rangle_0=
{UN_Ff_d\over (2\Delta)^2}[(v_yq_y)^2-i\omega\Gamma_{ph}].
\end{equation}
Here $\Gamma_{ph}$ is the phason damping rate and is given by
\begin{eqnarray}
\Gamma_{ph}(T\rightarrow T_c)=&2\Gamma_B\nonumber\\
\Gamma_{ph}(T\rightarrow 0)=&{8u_0^2T\Gamma_B\over
3\alpha^{4/3}G}e^{-G/T}.
\end{eqnarray}
The phason damping freezes out for low temperature, and
approaches $2\Gamma_B$ at $T_c$. Note the discontinuity in
$\Gamma_{ph}$ at $T_c$ (approaching from above $\Gamma_{ph}
\approx 2\pi^3T/7\zeta(3)$), which is the consequence of the
finite order parameter $\Delta$ below $T_c$ exceeding almost
immediately the energy scale set by $\omega$ and $vq$.

We are now able to write down the total correlation function
for this sound wave in the unpinned case. Using Eqs.(11),(18),
(24) and (25) we obtain
\begin{eqnarray}
\langle[\tau_{xy},\tau_{xy}]\rangle=&N_F(mv_Fv_y)^2
\times\nonumber\\
\times&\left [
{i\tilde\Gamma_{qp}(1-\tilde f)\over \omega+i\tilde
\Gamma_{qp}}+{(v_yq_y)^2f_d\over(v_yq_y)^2-i\omega\Gamma_{ph}}
\right ].
\end{eqnarray}
The above equation means that the collective contribution
(the second term on the right hand side) due
to the moving condensate recovers some of the screening of the
ion motion lost because of the decrease in the number of quasi
particles. In fact, at zero temperature it overcompensates
somewhat, since $f_d>\tilde f$. Therefore we expect the
electromechanical effect for the transverse sound polarized
in the chain direction only. According to Eq.(27), close to
$T_c$ the electromechanical effect on the sound velocity should
be much smaller than the temperature effect, while at low
temperatures the softening is somewhat bigger than the hardening
was upon cooling. Although the collective contribution in Eq.(27)
does have a frequency dependence, it does not appear to describe
the suppresion of the electromechanical effect when the
frequency is increased\cite{Xi3}. This is rather puzzling, although
the above results for sound propagation may not translate
literally to the flexural experiment.

\section{Conclusions}

We have derived for the first time the appropriate stress tensor for
quasi-one dimensional electron systems.
Under conditions of a sound velocity measurement ($lq\ll 1$) the
long-range Coulomb interaction has a simple role to suppress the
$s$-wave channel as in the theory of electronic Raman scattering.
In highly anisotropic systems like Bechgaard salts the increase of the
sound velocity in the density wave state is two orders of magnitude
smaller for longitudinal sound waves than for transverse ones. We also
find, that a sound wave with polarization perpendicular to the chain
direction can not couple to the phason, because the density wave
condensate can only move parallel to the chains. The coupling of the
longitudinal sound propagating in the chain direction is screened away
by the Coulomb interaction, which leaves us only the transverse
sound wave propagating perpendicular to the chains as the one which
does couple to the phason, and shows the electromechanical effect.

Recently Britel {\it et.al.} measured the elastic constant
$c_{44}$
of (Ta$_{1-x}$Nb$_x$Se$_4$)$_2$I at 15MHz in the geometry
${\bf u}\parallel {\bf x}$\cite{Mon}, and found a relative
reduction of order $10^{-4}$ in the presence of an electric field
approximately $10E_T$. This seems to be consistent with our
analysis, although the observed effect appears to be a little
too small.

\section*{Acknowledgments}

This publication is sponsored by the U.S.-Hungarian Science and Technology
Joint Fund in cooperation with the National Science Foundation and the
Hungarian Academy of Sciences under Project No. 264/92a, which enabled one
of us (A.V.) to enjoy the hospitality of USC. This work is also supported
in part by the National Science Foundation under Grant No. DMR92-18317 and
by the Hungarian National Research Fund under Grants No. OTKA15552 and
T4473.

\section*{Appendix}

We evaluate here the density correlation function $\langle[n,n]
\rangle_0^{no vertex}$ (Eqs.(12)-(14) in the text). We start with the
corresponding thermal product (See \cite{VM1}):
\begin{equation}
\langle[n,n]\rangle_0^{no vertex}=N_F[1-\pi T\sum_n F(iu_n,iu_n^\prime)],
\end{equation}
where $N_F$ is the density of  states, and $u_n$ and $u_n^\prime$ are
related to the Matsubara frequencies $\omega_n$ and $\omega_n^\prime=
\omega_{n-\nu}$ by
\begin{equation}
\omega_n/\Delta=u_n[1-\alpha(u_n^2+1)^{-1/2}],
\end{equation}
with $\Delta$ the order parameter, $\alpha=\Gamma/\Delta$ and
$\Gamma=\Gamma_F+\Gamma_B/2$ a combination of the impurity forward
and backscattering rates. Neglecting vertex corrections in the
relevant formulas in \cite{VM1} leads to:
\begin{equation}
F(u,u^\prime)={1+{1+uu^\prime\over (1-u^2)^{1/2}(1-u^{\prime 2})
^{1/2}}\over \Delta[(1-u^2)^{1/2}+(1-u^{\prime 2})^{1/2}]},
\end{equation}
where $u$ and $u^\prime$ are analytic continuations of $iu_n$ and
$iu_n^\prime$, and in the absence of the diffusion pole the
wavenumber ${\bf q}$ was already set to zero.

While evaluating the correlation function we follow the standard
method\cite{Zit}. Expanding up to linear order in $\omega$ we obtain
\begin{equation}
\langle[n,n]\rangle_0^{no vertex}=N_F(1-\tilde f+i\omega I),
\end{equation}
where
\begin{equation}
\tilde f={\pi T\over\Delta}\sum_n(u_n^2+1)^{-3/2},
\end{equation}
and
\begin{eqnarray}
I=&{1\over 2\Delta}\int_G^\infty{dE\over 2T}\cosh^{-2}\left ({E\over
2T}\right )\times\nonumber\\
\times&\left [{h^\prime\over {\rm Re}(1-u^2)^{1/2}}-{\rm Re}
(1-u^2)^{-3/2}\right ].
\end{eqnarray}
Here $h^\prime=(1/2)[1+(|u|^2+1)/|u^2-1|]$, and $G=\Delta u_0^3$ is
the gap with $u_0^2=1-\alpha^{2/3}$. The above equations can be
evaluated in two limiting cases, with temperature close to $T_c$ and
close to zero, and can be brought to the form of Eqs.(12)-(14) in
the text.

\end{document}